\documentclass[12pt]{article}
\usepackage{graphicx}
\usepackage{url}

\newcommand\pubnumber{SLAC--PUB--16768}
\newcommand\pubdate{November, 2016}

\def\SLAC{SLAC,
    Stanford University, Menlo Park, California 94025 USA}
\def\doeack{\footnote{Work supported by the US Department of Energy,
                     contract DE--AC02--76SF00515.}}

\def\Title#1{\begin{center} {\Large #1 } \end{center}}
\def\Author#1{\begin{center}{ \sc #1} \end{center}}
\def\Address#1{\begin{center}{ \it #1} \end{center}}

\def\submit#1{\begin{center}Submitted to {\sl #1} \end{center}}
\newcommand\pubblock{\rightline{\begin{tabular}{l} \pubnumber\\
         \pubdate \end{tabular}}}
\newenvironment{Abstract}{\begin{quotation} \begin{center}
                       ABSTRACT
     \end{center}\bigskip  }{\end{quotation}}

\def\submit#1{\begin{center}Submitted to {\sl #1} \end{center}}
\def\Acknowledgements{\bigskip  \bigskip \begin{center} \begin{large}
             \bf ACKNOWLEDGEMENTS \end{large}\end{center}}

\def\beq{\begin{equation}}
\def\eeq#1{\label{#1}\end{equation}}
\def\eeqn{\end{equation}}

\newenvironment{Eqnarray}%
   {\arraycolsep 0.14em\begin{eqnarray}}{\end{eqnarray}}
\def\beqa{\begin{Eqnarray}}
\def\eeqa#1{\label{#1}\end{Eqnarray}}
\def\eeqan{\end{Eqnarray}}
\def\CR{\nonumber \\ }

\def\leqn#1{(\ref{#1})}

\let\bar=\overbar

\def\lsim{\mathrel{\raise.3ex\hbox{$<$\kern-.75em\lower1ex\hbox{$\sim$}}}}
\def\gsim{\mathrel{\raise.3ex\hbox{$>$\kern-.75em\lower1ex\hbox{$\sim$}}}}

\def\half{\frac{1}{2}}

\def\del{\partial}
\def\Dslash{\not{\hbox{\kern-4pt $D$}}}
\def\dslash{\not{\hbox{\kern-2pt $\del$}}}

\def\msb{{\bar{\scriptsize M \kern -1pt S}}}

\def\drb{{\bar{\scriptsize D \kern -1pt R}}}

\makeatletter
\def\section{\@startsection{section}{0}{\z@}{5.5ex plus .5ex minus
 1.5ex}{2.3ex plus .2ex}{\large\bf}}
\def\subsection{\@startsection{subsection}{1}{\z@}{3.5ex plus .5ex minus
 1.5ex}{1.3ex plus .2ex}{\normalsize\bf}}
\def\subsubsection{\@startsection{subsubsection}{2}{\z@}{-3.5ex plus
-1ex minus  -.2ex}{2.3ex plus .2ex}{\normalsize\sl}}

\renewcommand{\@makecaption}[2]{%
   \vskip 10pt
   \setbox\@tempboxa\hbox{\small #1: #2}
   \ifdim \wd\@tempboxa >\hsize     
       \small #1: #2\par          
     \else                        
       \hbox to\hsize{\hfil\box\@tempboxa\hfil}
   \fi}

 \def\citenum#1{{\def\@cite##1##2{##1}\cite{#1}}}
 
\newcount\@tempcntc
\def\@citex[#1]#2{\if@filesw\immediate\write\@auxout{\string\citation{#2}}\fi
  \@tempcnta\z@\@tempcntb\m@ne\def\@citea{}\@cite{\@for\@citeb:=#2\do
    {\@ifundefined
       {b@\@citeb}{\@citeo\@tempcntb\m@ne\@citea\def\@citea{,}{\bf ?}\@warning
       {Citation `\@citeb' on page \thepage \space undefined}}%
    {\setbox\z@\hbox{\global\@tempcntc0\csname b@\@citeb\endcsname\relax}%
     \ifnum\@tempcntc=\z@ \@citeo\@tempcntb\m@ne
       \@citea\def\@citea{,}\hbox{\csname b@\@citeb\endcsname}%
     \else
      \advance\@tempcntb\@ne
      \ifnum\@tempcntb=\@tempcntc
      \else\advance\@tempcntb\m@ne\@citeo
      \@tempcnta\@tempcntc\@tempcntb\@tempcntc\fi\fi}}\@citeo}{#1}}
\def\@citeo{\ifnum\@tempcnta>\@tempcntb\else\@citea\def\@citea{,}%
  \ifnum\@tempcnta=\@tempcntb\the\@tempcnta\else
  {\advance\@tempcnta\@ne\ifnum\@tempcnta=\@tempcntb \else\def\@citea{--}\fi
    \advance\@tempcnta\m@ne\the\@tempcnta\@citea\the\@tempcntb}\fi\fi}
\makeatother

\begin{document}
\begin{titlepage}
\pubblock

\vfill
\Title{Revision of the LHCb Limit on Majorana Neutrinos}
\vfill
\Author{Brian Shuve and Michael E. Peskin\doeack}
\Address{\SLAC}
\vfill
\begin{Abstract}
We revisit the recent limits from LHCb on a Majorana neutrino $N$  in the
mass 
range 250--5000 MeV~\cite{Aaij:2014aba}. These limits are among the best currently
available, and they will be improved soon by the addition of data from
Run 2 of the LHC.    LHCb presented a model-independent constraint on
the rate of like-sign leptonic decays, and then derived  a constraint on the mixing
angle $V_{\mu 4}$  based on a theoretical model for the $B$ decay width
to $N$  and the $N$  lifetime.
 The model used is unfortunately unsound.  We revise
the conclusions of the paper based on a decay model similar to the one
used for the $\tau$  lepton and provide formulae useful for future analyses.
\end{Abstract}
\vfill
\submit{Physical Review {\bf D}}
\vfill

\newpage
\tableofcontents
\end{titlepage}

\def\thefootnote{\fnsymbol{footnote}}
\setcounter{footnote}{0}

\section{Introduction}

One of the important questions in neutrino physics is that of the
existence of massive sterile neutrinos.   Neutrinos with zero charges
under $SU(2)\times U(1)$ could potentially exist at any of a
number of different mass scales.  For eV-scale masses, massive neutrinos
are invoked to explain anomalies in 
short-baseline neutrino 
experiments~\cite{Aguilar:2001ty,AguilarArevalo:2008rc,AguilarArevalo:2010wv}.
For masses of $10^{9}$--$10^{12}$~GeV,   massive neutrinos  are invoked
in the seesaw mechanism that leads to small masses for the known
neutrinos even in the presence of large Yukawa 
couplings~\cite{Minkowski:1977sc,GellMann:1980vs,Yanagida}. 
However, massive sterile neutrinos could potentially
exist at any mass scale, provided they are 
mixed with the known
leptons sufficiently weakly.

For a massive neutrino in the GeV mass range, the Belle and LHCb
experiments have recently presented new limits based on the possible
appearance of such neutrinos in $B$ meson decays.  In the processes
studied, the massive neutrino $N$ would be a $B$ meson decay product with an
observable lifetime that decays to  the final state $\pi^\pm\ell^\mp$.
Belle has searched for the production of $N$ in $B \to D, D^* + \ell
N$,  as well as in $B \to X \ell N$ in an analysis valid for $m_N > 2$~GeV,
where $X$ includes a series of light-quark
mesons~\cite{Liventsev:2013zz}.   The originally published Belle analysis over-estimated their 
limit due to an improper treatment of the $N$ lifetime distribution.
A recent update of the analysis  corrects this point~\cite{Liventsev_PC}.

The LHCb experiment has published the 
results of a dedicated search for a massive Majorana neutrino 
in the lepton-number-violating decay
\beq
              B^- \to \mu^- N \ , \qquad   N \to \pi^+\ \mu^- 
\eeq{thedecay}
and its charge-conjugate process \cite{Aaij:2014aba}. 
Searches are performed for both prompt and displaced $N$ decays.
The requirement of like-sign leptons and the
requirement that the three final charged particle momenta sum to the
$B$ mass reduce the background in this hadron collider experiment to
about 70 events, scattered in $m_N$ over the window from 0.5 to 5 GeV.
The limit obtained on $|V_{\mu 4}|^2$ is stronger than the Belle limit
at high values of the $N$ mass.   For lower mass values, it is likely
be competitive in the future, with  a  large new data sample from the
LHC run 2 and the prospect of including
events with $e^\pm$ and other production and decay modes.

The analysis in \cite{Aaij:2014aba} provides  model-independent limits
on the rate 
of like-sign leptonic $B$ decays and then interprets these limits
 in terms of $|V_{\mu 4}|^2$  based on a theoretical model. 
Unfortunately, the model described in  \cite{Aaij:2014aba}  for the rates of the
relevant $B^-$ and $N$ decays is unsound, with dependences 
on $m_N$ that are difficult to defend theoretically. Replacing these
expressions 
with more correct ones significantly changes the quoted limits.
In view of the promise of this experiment to probe for Majorana 
neutrinos at a very sensitive level with future data expected from the 
LHC, we felt that it would be useful to present an improved theory of 
these decays and to compute more correct limits on $|V_{\mu 4}|^2$
based on the results of \cite{Aaij:2014aba}.

\section{Decay rates involving $N$}

The analysis of  \cite{Aaij:2014aba} proceeds in two stages.  First,
model-independent limits are placed on the product of branching ratios  
$\mathrm{BR}(B^-
\to \mu^- N) \cdot \mathrm{BR}(N\to \pi^+\mu^-)$
as a function of the mass $m_N$ and decay rate $\Gamma_N$.  These constraints are then reinterpreted as 
limits on the sterile neutrino mixing angle $|V_{\mu 4}|^2$ using theoretical expressions
that relate $\Gamma_N$ and $m_N$.   The efficiency for detecting the
events as a function of the lifetime of $N$ plays a non-trivial role. 

The decay rates of a massive Majorana neutrino into a variety of
relevant final states have been computed by Gorbunov and
Shaposhnikov~\cite{Gorbunov:2007ak},
and Atre, Han, Pascoli, and Zhang~\cite{Atre:2009rg}.  In the
following expressions, 
we neglect final-state masses for all particles except the $N$, which is sufficient to demonstrate the parametric dependence of the decay rates.   The partial
widths relevant to this analysis are:
\beqa
 \Gamma(B^- \to \mu^- N) &=&   {G_F^2\over 8\pi} m_N^2 f_B^2 m_B
 |V_{\mu 4}|^2 |V_{ub}|^2
\bigl(1 - {m_N^2\over m_B^2}\bigr)^2 \ ,  \CR
   \Gamma(N \to \pi^+\mu^-) &=& {G_F^2\over 16\pi} m_N^3 f_\pi^2
      |V_{\mu 4}|^2 |V_{ud}|^2 \ ,
 \eeqa{BGammas}
where $f_i$ is the leptonic decay constant for pseudoscalar $i$, $G_F$
is the Fermi constant, and 
$V_{qq'}$ is the relevant CKM matrix entry. This gives
\beqa
 \mathrm{BR}(B^-
\to \mu^- N)\cdot  \mathrm{BR}(N\to \pi^+\mu^-)  & & \CR & &  \hskip
-1.2in = \tau_B\tau_N
\cdot {G_F^4 f_B^2 f_\pi^2 m_B
  m_N^5 \over 128 \pi^2} |V_{\mu 4}|^4|V_{ub}|^2 |V_{ud}|^2 
 \bigl(1 - {m_N^2\over m_B^2}\bigr)^2 \ , 
\eeqa{Bfinal}
where $\tau_B$ and $\tau_N$ are the lifetimes of $B$ and
$N$. This result is similar to eq.~(1) of \cite{Aaij:2014aba}, which
is based on eq. (3.30) of \cite{Atre:2009rg}.  It differs, however, by
a factor $m_N^4/m_B^4$, which gives a substantially smaller signal
rate at low values of $m_N$.   Of this, one factor of $m_N^2/m_B^2$
comes from the helicity suppression of the first process in
\leqn{BGammas}, while the other factor of $m_N^2/m_B^2$ comes from the
reduced phase space for the on-shell $N$ decay.   We note that the
result in \cite{Atre:2009rg} is given only as an estimate for general
lepton number violating rare decays.

The $B$ lifetime is known, but the lifetime of $N$ must be computed
from theory.   The leptonic partial widths are straightforwardly
computed.  For the hadronic decays, it has been understood for a
long time that the total hadronic rate for weak interaction decays of a lepton  is well estimated
by the QCD decay to quark-antiquark pairs even for a mass as low as
$m_\tau = 1.78$~GeV~\cite{Lam:1977cu,Braaten:1988hc,Narison:1988ni}.
This is now the basis of our precision understanding of inclusive
$\tau$ decays~\cite{Pich:2013lsa}.   The same formalism should apply
to compute the lifetime of the $N$ when the $N$ has a mass comparable
to or higher than  $m_\tau$.   Note that, at the same time that  we use this QCD
estimate to compute the total decay rate of the $N$, we must still use \leqn{BGammas} to
compute the rate for the exclusive decay to 
$\pi^+\mu^-$.

 For Majorana $N$, the leptonic charged-current decay rate is
\beq
    \Gamma(N\to \ell_i^- \ell_j^+ \nu_j) = {G_F^2 m_N^5\over
    192\pi^3}|V_{i4}|^2 \ , 
\eeq{diffl}
 In this equation, $i, j = 1,2,3$ run over the three generations,
with $i \neq j$.
Also, here and in all cases below, we   must  add an equal rate for
the separate decay to the charge-conjugate channel (here, $N\to \ell_i^+\ell_j^- \bar\nu_j$).  The
hadronic charged-current decay rate  is 
\beq
    \Gamma(N\to \ell_i^- u_j \bar d_j) = 3\, (1 + {\alpha_s\over \pi }
    + \cdots) \ {G_F^2 m_N^5\over
    192\pi^3}|V_{i4}|^2 \ . 
\eeq{hadronic}
Here,  $u_j= (u,c)$, $d_j = (d,s)$; the third generation is
kinematically inaccessible.  We can ignore CKM mixing in computing
the total rate  in the limit of massless 1st and 2nd generation quarks.

  The
neutral current decay rate to a quark or lepton $f$ depends on the
electric charge $Q_f$ and the left- and right-handed $Z$ charges
\beq
     Q_{ZfL} = \pm \half - Q_f \sin^2\theta_w  \qquad   Q_{ZfR} = -
     Q_f \sin^2\theta_w  \ ,
\eeqn
where the $+$ applies to up-type quarks and neutrinos, and the $-$ applies to down-type quarks and charged leptons. Let 
\beq
   S_f =Q_{ZfL}^2 + Q_{ZfR}^2  =   {1\over 4} - |Q_f|\sin^2\theta_w +
 2  Q_f^2 \sin^4\theta_w \ .
\eeqn
Then, for $f \neq \nu_i$,     
\beq
    \Gamma(N\to \nu_i^- f\bar f) = {G_F^2 m_N^5\over
    192\pi^3}|V_{i4}|^2 \ S_f \ ,
\eeq{neutralcurrent}
The decay rate for $N\rightarrow \nu_i \nu_i \bar\nu_i$ is 2 times the
rate for
$N\rightarrow \nu_i \nu_j\bar\nu_j$, $i\neq j$.
Finally, for the charged lepton decay with $i = j$,  
\beq
    \Gamma(N\to \nu_i^- \ell^+_i\ell^-_j) = {G_F^2 m_N^5\over
    192\pi^3}|V_{i4}|^2 \ \bigl( {1\over 4} +\sin^2\theta_w +
 2  \sin^4\theta_w) \ .
\eeq{samel}

 We now evaluate these formulae for the case $|V\mu 4|^2 \neq 0$,
  $|V_{i4}|^2 = 0 $ for $i \neq \mu$,  which is constrained by the LHCb measurement.
  The charged-current widths to $e\mu\nu$, $ee\nu$ and $\mu\mu\nu$ are,
  respectively
\beq
     \Gamma_{\ell\ell \nu}  =  {G_F^2 m_N^5\over
    96\pi^3}\, |V_{\mu 4}|^2\, (1 + 0.13 + 0.59) \ .
\eeq{lldecay}
The charged-current and neutral-current quark contributions, treating
$u,d,s,c$ as massless,  give 
\beq
    \Gamma_{qq}  =  {G_F^2 m_N^5\over
    96\pi^3}\,|V_{\mu 4}|^2\, \cdot (8.24) \ .
\eeqn
The neutral-current widths to 3 neutrinos gives
\beq
\Gamma_N \approx {G_F^2 m_N^5\over
    96\pi^3}\,|V_{i4}|^2\, \cdot ( 1.00)   \ .
\eeqn
The sum of these gives the total width,
\beq
\Gamma_N \approx {G_F^2 m_N^5\over
    96\pi^3}\, |V_{i4}|^2\, \cdot ( 10.95)  \ .
\eeq{totalG}
The last numerical factor becomes (12.08) when we include decays to
$\tau$ while ignoring the $\tau$ mass.
The dependence on the mass of the $N$ is always  $m_N^5$.  
This contrasts with eq.~(2) of \cite{Aaij:2014aba}, which has the same
structure for the leptonic decay width but behaves as $m_N^8$ for the
hadronic part of the decay width.   Our  formula \leqn{totalG} is
considerably smaller at large values of $m_N$ than the width employed in the LHCb interpretation.

For small values of the mass of the $N$,  the QCD estimate for the total
decay rate will break down, and the decay rate must be computed as the
sum of exclusive decays to charged and neutral hadrons.  We have
computed the total decay rate for $N$ in this way, summing over 
charged and neutral current hadronic  decays to  $\nu_i \pi$
and $\nu_i \rho$.  This approach gives  values
within 15\% of the QCD formula in the region  $1.0~\mathrm{GeV} < m_N < 1.5$~GeV. 
  As we  discuss below, the transition
makes little difference to the final result since, in this mass range,
 the limit on $|V_{\mu 4}|^2$ is 
 insensitive to the $N$ lifetime.

\section{Results}

\begin{figure}
\begin{center}
\includegraphics[width=0.750\hsize]{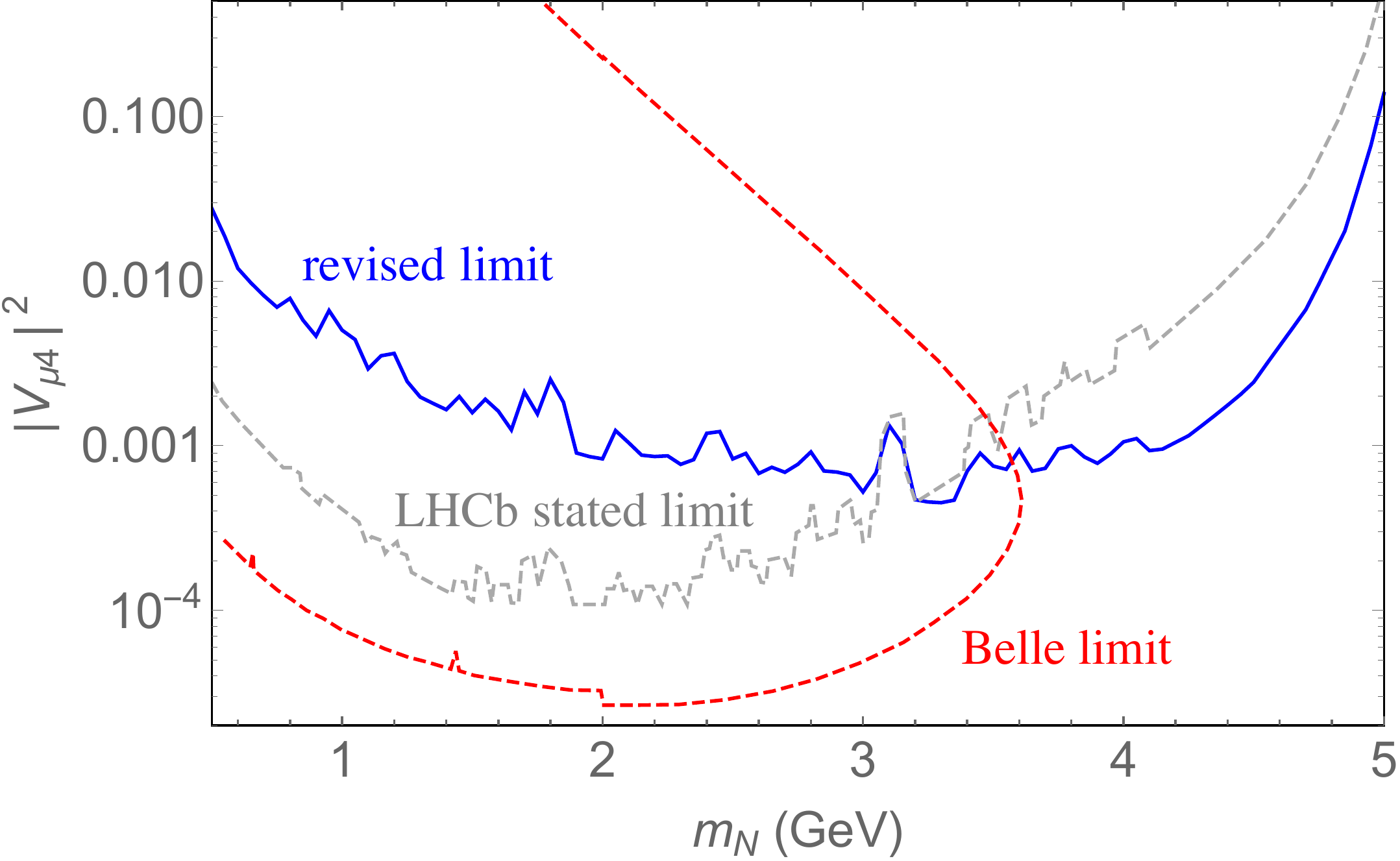}
\caption{Upper limit on $|V_{\mu 4}|^2$ at 95\% confidence level from
  the LHCb experiment.  The
  dashed line shows the limit from \cite{Aaij:2014aba}.   The solid
  line shows the limit that would be extracted using the decay width formulae in
  this paper. For comparison, the lower dotted line shows the recently revised  limit
  from Belle \cite{Liventsev:2013zz,Liventsev_PC}. All three limit
  curves are constructed with the assumption  $V_{e4} = V_{\tau4} = 0$. }
\label{fig:bound}
\end{center}
\end{figure}

Using the above results, we  convert the model-independent LHCb limits
on the rate of like-sign lepton decays of $B^-$ into limits on
$|V_{\mu 4}|^2$ as a function of $m_N$,
 assuming $|V_{i4}|^2=0$ for $i=e,\tau$.  In our analysis, we compute the $N$ production 
rate as in (\ref{Bfinal}), while we compute the total width by modeling the hadronic decays with the quark model for $m_N \ge1.5$ GeV as in (\ref{totalG}),
and with a sum over exclusive hadronic decay modes for $m_N < 1.5$ GeV. We also restore 
all final-state masses in the expressions for the $B^-$ and $N$
decay 
rates~\cite{massive}, which were neglected above to simply the analytic expressions.    

Ref.~\cite{Aaij:2014aba}
provides limits on the product of branching ratios $ \mathrm{BR}(B^-
\to \mu^- N)\cdot  \mathrm{BR}(N\to \pi^+\mu^-)$ for certain values of 
the lifetime $\tau_N$; we interpolate between the given values, and
for 
very large $\tau_N$, where the mean decay length is far outside the
detector, 
we use a decay acceptance inversely proportional to the lifetime. 
 For each value of $m_N$, we iteratively scan through values of
 $|V_{\mu 4}|^2$, 
determining the mixing angle for which the computed $N$ production
rate is
 equal to the LHC constraint on $ \mathrm{BR}(B^-
\to \mu^- N)\cdot  \mathrm{BR}(N\to \pi^+\mu^-)$ for the lifetime corresponding to $|V_{\mu 4}|^2$.   To evaluate \leqn{Bfinal}, we use the same values as LHCb to facilitate comparison:~$f_B =0.19$ GeV, $f_\pi =0.131$ GeV,
$|V_{ub}| = 0.004$, $|V_{ud}|=0.9738$, $M_B=5.279$ GeV, $\tau_B=1.671$ ps.  The uncertainties in these quantities have only a
small effect on the quoted limits.

The effect of the updated analysis, shown in Fig.~\ref{fig:bound}, is
substantial.   To understand this, first note that $\mathrm{BR}(N\to
\pi^+\mu^-)$ includes the factor  $\Gamma_N^{-1}$ and so is linearly
proportional to $\tau_N$.    With this in mind,  the
differences between our result and that  of \cite{Aaij:2014aba} come
from two effects: At low values of $m_N$ (below 2~GeV), the change in
eq.~\leqn{Bfinal} 
  leads to a 
substantially smaller event rate at low
values of $m_N$ (below 2 GeV).   In this region, the limit on the
mixing angle is 
 largely insensitive to the lifetime $\tau_N$.  The reason for this is
 that the decay length is sufficiently long that the decay acceptance
 is 
inversely proportional to $\tau_N$, cancelling the factor of $\tau_N$
from   $\mathrm{BR}(N\rightarrow\mu^+\pi^-)$.  At high 
values of $m_N$ (above 3 GeV),  the updated $\tau_N$ is significantly
larger than before, leading to a larger
$\mathrm{BR}(N\rightarrow\mu^+\pi^-)$. In this region, most $N$ decays occur
inside the detector and so this change is mainly reflected in  
a larger signal rate predicted by theory 
and, consequently, a stronger limit.

We look forward to a substantial improvement in the limits
 on $|V_{\mu 4}|^2$ from LHCb  using the large data sets that will be
available from the LHC Run 2 and beyond. 

\newpage

\Acknowledgements

We are grateful to Sheldon Stone for his encouragement and for his
help in understanding the LHCb analysis.  We thank  Dmitri
Liventsev for discussions of the Belle analysis and for providing us the
updated Belle limits.
This work was supported by the U.S. Department 
of Energy under contract DE--AC02--76SF00515.

\end{document}